\documentstyle[12pt]{article}
\begin{document}
\setlength{\baselineskip}{0.25in}

\newcommand{\beq}{\begin{equation}}
\newcommand{\eeq}{\end{equation}}
\newcommand{\beqa}{\begin{eqnarray}}
\newcommand{\eeqa}{\end{eqnarray}}
\newcommand{\ls}{L\"{u}scher-Schechter\,\,}
\newcommand{\lsim}{\begin{array}{c}\,\sim\vspace{-21pt}\\<\end{array}}
\newcommand{\gsim}{\begin{array}{c}\sim\vspace{-21pt}\\>\end{array}}
\newcommand{\bfk}{{\bf k}}
\newcommand{\ak} {a(\bfk )}
\newcommand{\aks}{a^*(\bfk )}
\newcommand{\bk} {b(\bfk )}
\newcommand{\bks}{b^*(\bfk )}
\newcommand{\once}{{8\pi^2\over g^2}}
\newcommand{\twice}{{16\pi^2\over g^2}}

\begin{titlepage}

{\hbox to\hsize{May 1993 \hfill JHU-TIPAC-930013}}

{\hbox to\hsize{ \hfill hep-ph/9305234 }}

\begin{center}
\vglue .06in
{\Large \bf Complex Time Solutions with Nontrivial Topology
and Multi Particle Scattering in  Yang-Mills Theory}

\vspace{1.5cm}

\begin{tabular}{c}
{\bf Thomas M. Gould\footnotemark[1]\footnotemark[2]\footnotemark[3]}
and {\bf Erich R. Poppitz\footnotemark[1]\footnotemark[4]}\\[.05in]
{\it Department of Physics and Astronomy}\\
{\it The Johns Hopkins University}\\
{\it Baltimore MD 21218 }\\[.15in]
\end{tabular}

{Abstract}\\[-.05in]

\footnotetext[1]{Supported in part by the National Science Foundation
under grant PHY-90-9619.}
\footnotetext[2]{Supported by the Texas National Research Lab Commission
under grant RGFY-93-292.}
\footnotetext[3]{gould@fermi.pha.jhu.edu}
\footnotetext[4]{poppitz@dirac.pha.jhu.edu}
$\bigtriangledown $
\end{center}

\begin{quotation}
A classical solution to the Yang-Mills theory is given
a new semiclassical interpretation in terms of particle scattering.
It solves the complex time boundary value problem,
which arises in the semiclassical approximation to a multi particle
transition probability in the one-instanton sector at fixed energy.
The imaginary part of the action of the solution on the complex time
contour and its topological charge obey the same relation as the
self-dual Euclidean configurations.
Hence the solution is relevant for the problem of tunneling with
fermion number violation in the electroweak theory.
It describes transitions from an initial state
with a smaller number of particles to a final state with
a larger number of particles.
The implications of these results for multi particle production in the
electroweak theory are also discussed.
\end{quotation}
\end{titlepage}

\newpage

An intriguing feature of the Yang-Mills gauge theory is the periodic
structure of its vacuum \cite{JAC,CDG}.
In the semiclassical approximation,
the topology of finite energy solutions leads to a classification
of all gauge-inequivalent vacua in the theory.
The discovery of this rich structure has had a profound impact on our
understanding of non-perturbative aspects of the theory,
notably a low energy phenomenon like the solution of the famous U(1)
problem in QCD \cite{THO}.
However,
the role of the vacuum in the dynamics of particle scattering
has not yet been as deeply understood.
This deficit in our understanding has been confronted in recent years
 with the study of so-called ``instanton-induced'' cross-sections
\cite{RIN,KRT,early}.

The simplest semiclassical estimate of the contribution of
the BPST \cite{BPST} instanton to a total inclusive
two particle cross-section
in electroweak theory implies
a result which grows exponentially with center-of-mass energy \cite{RIN,KRT}.
It has been shown that the instanton is the basis of
a systematic perturbative expansion of the final state radiative
corrections to the cross-section \cite{KRT},
which determines its leading semiclassical behavior,
neglecting initial state radiative corrections,
\beq
\label{eq:holygrail}
\sigma_{tot} (x) \; \sim \;
\exp{\left[\, \twice F(x) \, + \, o(\alpha^0)\,\right]}
\eeq
as an expansion in powers of a small parameter $x\equiv E/E_0$,
the ratio of the center-of-mass energy $E$ and a mass scale of
order the electroweak sphaleron mass,
$E_0 \simeq M_w/\alpha_w \simeq 10\: TeV$.
The so-called ``Holy Grail function'', $F(x)$, is approximately
$-1$ for small $x$, reflecting
the severe\, 'tHooft suppression factor,
$\exp{-\twice} \simeq 10^{-127}$, due to the large instanton action.
The fact that the Holy Grail function is an increasing function of $x$
for small $x$ has led many to speculate about the possibility of overcoming
the severe exponential suppression factor at energies of order $E_0 $.
The possibility of strong multi-particle scattering in electroweak theory
at multi-TeV  energies has led to an enormous effort to understand the behavior
of multi-particle cross-sections in the sphaleron
energy regime \cite{early}.

Most of these studies, however,
are based on semiclassical expansions around
configurations which are not influenced by external sources.
Indeed,
the instantons obey vacuum boundary conditions,
and as such are relevant to this problem only in the approximation in
which external sources are neglected.
While the final state corrections can be accounted for in the perturbative
expansion in $x$,
the initial state corrections are more subtle.
These involve radiative corrections to hard particles which are not
{\it a priori} calculable semiclassically.
However,
there have been some indications \cite{MUE}
that the contributions to the leading semiclassical result due
to corrections involving hard initial legs may also be calculable
in a semiclassical manner.
It may then be possible to calculate the entire leading order
semiclassical exponent in a saddle point approximation.
What is needed  is a new technique which accounts for external sources
to make the semiclassical behavior of the total cross-section manifest.

A strategy for out-flanking the problem of initial state corrections
was recently proposed by Rubakov and Tinyakov \cite{RT}.
The basic idea is to consider transitions from states of a fixed large
number of particles, say $N_{\rm in} = \nu/g^2$.
The instanton-like transition probability
from a multi-particle initial state is then calculable semiclassically,
in the limit $g \rightarrow 0$ with
$\nu$ fixed.
Its leading semiclassical behavior is  determined by the solution to a boundary
value problem.
The boundary conditions imposed at initial and final times
correctly account for the energy transfer from the initial multiparticle state
to the final multiparticle state.
The leading semiclassical behavior of the $N_{\rm in}$ particle transition
probability
has a form similar to (\ref{eq:holygrail}).
It is a rigorous upper bound on the inclusive two-particle
cross section
and is related to a lower bound under less rigorous assumptions \cite{KRT3}.
Since the $N_{\rm in}$ particle transition probability
contains all initial state corrections for the
$N_{\rm in} = \nu/g^2$ particle transition,
it is hoped that it reproduces the leading semiclassical behavior
of a two particle transition when $\nu$ is small,
including initial and final state corrections.
Indications from explicit calculations of initial and final
state corrections are that the limit of $\nu\rightarrow 0$ is smooth
\cite{MUE},
so that the contribution to a semiclassical transition probability
from the solution of the boundary value problem contains the initial
and final state corrections.
The boundary value problem posed in this way also holds the promise of
being amenable in principle to numerical computation of multiparticle
transitions.
It would now be useful to have some analytical examples to guide future
efforts in this direction \cite{RSTnew,KYA,GP}.

The use of  a Minkowski or Euclidean time contour for the semiclassical
calculation of
transition amplitudes in the one-instanton sector is too restrictive.
Recall that computing tunneling contributions  to fixed-energy (i.e.
time-independent)
Green functions in quantum mechanics can be performed in the WKB approximation
only on a complex time contour, chosen to lie in Minkowski directions at early
and late times,
with a period of Euclidean evolution inserted at an intermediate time.
They give the dominant WKB-contribution to classically forbidden processes.
In the present case of quantum field theory,
we will similarly be interested only in time-independent transition
probabilities.
\begin{picture}(300,200)(-60,0)
\put (50,100){\vector(1,0){200}}\put(150,40){\vector(0,1){130}}
\thicklines\put(70,140){\line(1,0){80}}\put(150,140){\line(0,-1){40}}
\put(150,100){\line(1,0){80}}\put(100,125){\circle*{5}}
\put(110,125){$\tau_{1,-1}$}\put(100,148){\circle*{5}}
\put(110,148){$\tau_{2,-1}$}\put(65,145){$C_T$}\put(157,135){$T$}
\put(155,165){Re~$\tau =$ Im~$t$}\put(240,87){Im~$\tau =$ Re~$t$}
\put (135,20){Fig. 1}
\end{picture}

In nonabelian gauge theories,
the transition amplitude between vacua with different topological number
is known to be maximized by instantons.
They and, in fact, any finite action Euclidean solution have vacuum asymptotics
at infinity.
For transitions involving many-particle initial and final states,
vacuum boundary conditions are clearly not the correct ones.
Considering solutions on a complex time contour, $C_T$, (fig.1) provides
a natural description of the initial and final states in Minkowski space
in terms of the free wave asymptotics of the solution at
$\vert\, {\rm Re}\, t\,\vert \rightarrow\infty$.

A few such solutions on a complex time contour have already been investigated.
The {\it periodic instanton} is a solution of the complex time boundary value
problem
which arises from the semiclassical approximation to the inclusive transition
probability
from all initial states at fixed energy, or a microcanonical distribution
\cite{KRT3}.
It has two turning points on the complex time contour.
It has been shown to determine the maximal probability for transition
in the one-instanton sector from states of fixed energy \cite{KRT3,RT}.
The periodic instanton in electroweak theory has so far been
constructed only in a low energy approximation,
and the resulting transition probability is determined in
a perturbative expansion similar to that in (\ref{eq:holygrail}).
It has been found to describe transitions between states of equal number of
particles
which is large in the semiclassical limit, $N_{\rm in} = N_{\rm
fin} \sim 1/g^2$.
So, this solution is irrelevant for describing $2 \rightarrow n$ scattering
processes at high energies,
though it does play a role in determining the rate of tunnelling,
and anomalous baryon number violation, at finite temperature \cite{HSU}.

A solution which describes transitions from a state of smaller number of
particles to a state with a larger number
of particles has also only been constructed in a low energy expansion
\cite{RT}.
Similarly,
it determines the maximum transition probability from states of
fixed energy and particle number.
However,
it remains to construct solutions which describe such processes in general.
This is a formidable task, requiring a solution of the Yang-Mills equations
with arbitrary boundary conditions on a complex time contour.
In this paper, we pursue more modest goals.
We consider the SO(4) conformally invariant Minkowski time
 solutions of L\"uscher and Schechter \cite{LUS},
analytically continued to a complex time contour (fig. 1).
 The solution in Minkowski-time describes an energy density which
evolves from early times as a thin collapsing spherical shell,
bounces at an intermediate time,
and expands outward again at late times.
As yet,
the role of these solutions in scattering problems has not been
fully developed \cite{FKS}.

Our aim is the calculation of many-particle transition amplitudes in the
one-instanton sector.
Therefore,
we will consider only a subclass of these solutions which have
integer topological charge on the complex time contour $C_T$.
Only the solutions with a turning point at say, $t=0$ for all $\vec{x}$,
have this property, as we will show.
The \ls solutions are real in Minkowski time.
The turning point condition assures that their analytic continuation to
the Euclidean time axis is real as well.
Note that in general the fields will be complex on the ${\rm Re}\, t<0$
part of the contour,
since $t=iT$ is {\it not} a turning point of the
solution\footnote{It is easy to show that the SO(4)-conformally invariant
solutions
can have at most one turning point.}.

L\"uscher and Schechter have shown that the most general
solution for which a $SO(4)$-conformal transformation
can be compensated by a global $SU(2)$-gauge transformation
is parameterized by a single function  $q(\phi)$,
where
$2\, r\cosh \phi = \left( 1 + r^2 + \tau^2\right) \cos w , \hspace{2mm}
r\sinh \phi = \tau\cos w $ and $ 2\, r\tan w = r^2 + \tau^2 -1$.
Its Euclidean action  is \cite{LUS}:
\beqa
\label{a2}
S & = &
 {i\over 4 g^2}\,\int d^4 x \,  F_{\mu\nu}^a F_{\mu\nu}^a \: = \:
i\, {12\pi \over g^2}\int^{\infty}_0 d r
\int_{-\infty}^{+\infty}d\tau \:
{\cos^4 w \over r^2} \,
\left[\,
{1 \over 2}\,\dot{q}^2 \, +\,{1 \over 2}\, \left(q^2 - 1\right)^2
\,\right] \, ,
\eeqa
where $\dot{q} \equiv {d \over d\phi} q(\phi)$, $\tau \, \equiv\,  it$
is the Euclidean time and $r\equiv \vert\vec{x}\vert$.

The topological charge in terms of the \ls Ansatz becomes :
\beqa
\label{q1}
Q & \equiv &
{1 \over 32\pi^2}\int d^4 x \: F_{\mu\nu}^a \tilde{F}_{\mu\nu}^a \: = \:
{1 \over 2\pi} \int\limits^{\infty}_0 dr
\int\limits_{-\infty}^{+\infty}d\tau \:
{\cos^4 w \over r^2}\,\dot{q}\, \left(\, 3q^2 \, -\, 3 \,\right) \, .
\eeqa

A solution with a turning point is easy to find
by considering the one-dimensional double-well problem, following from
(\ref{a2}).
It represents oscillatory motion in the well between $q=1$ and $q=-1$
of the potential $V(q)\, = \, {1 \over 2}\, \left(q^2 - 1\right)^2$ .
The turning point condition\footnote{One can  verify,
using the explicit formulae relating $q$ to gauge potentials \cite{GP},
that the condition $\dot{q}(\phi = 0)=0$ corresponds to a turning point
of the gauge potentials at $\tau = 0$ for all $\vec{x}$.}
 at $\phi = 0$
 leaves one free parameter:
the ``energy'' $\epsilon$ ($\epsilon < 1/2$),
or equivalently the initial coordinate, $q_- = \sqrt{1-\sqrt{2\epsilon}}$,
of the particle in the well.
This solution is explicitly given in terms of the Jacobian elliptic
sine:
\beq
\label{soltn}
q\left(\,\phi (r, \tau)\,\right) \: = \:
q_{-}\, {\rm sn} \left(\, q_+\phi (r, \tau) \, + \, K \, , \, k\,\right) \, .
\eeq

Now,
the imaginary part of $S_{C_T}$ is the quantity entering the WKB-exponent
of a transition probability dominated by this solution.
Since the solution is real on the Minkowski time axis,
the contribution to the action from the real time axis is purely real.
So, the residue at the singularity between the complex time contour
and the Minkowski time axis alone determines ${\rm Im}\, S_{C_T}$ :
\vspace{1mm}
\beq
\label{ims}
{\rm Im}\, S_{C_T} \: = \:
 -{24\pi^2 \over g^2}\int^{\infty}_0 dr \:
{\rm Im} \sum_{nm} {\rm Res}\,
\left\{\, {\cos^4 w \over r^2}\,
\left[\,
{1 \over 2}\,\dot{q}^2 \, +\,{1 \over 2}\, \left(q^2 - 1\right)^2
\,\right]
\,\right\}
\rule[-3mm]{0.2mm}{9mm}_{\,\tau_{nm}(r)} \, .
\eeq

Consider now the topological charge $Q$ (\ref{q1}) on the closed contour
$C_T + C_M$ where $C_M$ runs along the Minkowski time axis.
Our solution (\ref{soltn}) is
an even function of $\phi$, therefore the integral for $Q$ on the Minkowski
time axis vanishes.
Thus,
$Q$ on $C_T$ is determined by the residues at the singularity  as well:
\vspace{1mm}
$$
Q \: = \:  3\, i\,\int^{\infty}_0 dr\:
\sum_{nm} \, {\rm Res}\,
\left\{\,
{\cos^4 w \over r^2}\, \dot{q} \, \left(q^2 - 1\right)
\,\right\}\rule[-3mm]{0.2mm}{9mm}_{\,\tau_{nm}(r)} \, .
$$

Explicit calculation yields for the imaginary part of the action and the
topological charge:
\beq
\label{action}
{\rm Im}\, S \: = N \: {8\pi^2 \over g^2} \, , \hspace{.5cm} Q \: = \: N,
\eeq
where $N$ is the number of singularity lines between the
complex time contour and the Minkowski time axis.
It should be stressed that the relation (\ref{action}) is
far from trivial on the complex time contour.
The usual arguments for establishing the
Bogomol'nyi bound  do not seem to hold here,
since the fields take complex values on the contour \cite{BOG}.
The turning point condition at $\tau =0$ is crucial for (\ref{action}) to hold.
As was shown in \cite{FKS},
the Minkowski time topological charge vanishes only for solutions
with a turning point.
The \ls\hspace{.2cm}solutions without a turning point have fractional
topological
charge on the contour $C_T$\footnote{The charge on the contour $C_T$ is
in this case the sum of an (integer) residue and a (fractional \cite{FKS})
Minkowski
time contour contribution.}.

This gauge field configuration gives the dominant contribution
to an inclusive transition probability from a fixed initial
state, in the saddle point approximation \cite{RSTnew}.
The initial state, and the most probable final state for transition from
this initial state, are characterized by the free-wave asymptotics
of the solution at the ends of the complex time contour $C_T$.
The total transition probability from an initial coherent state,
$ \vert\,\{ \ak \}\,\rangle $,
projected onto  fixed center-of-mass energy $E$, is:
\beq
\label{eq:probability}
\sigma_E \left( \{ \ak \} \right) \: = \:
\sum_{f} \,\vert\,
\langle\, f \,\vert\,\hat{S}\, P_Q\, P_E\, \vert\, \{ \ak \} \,\rangle\,
\vert^{\, 2}   \, .
\eeq
$P_E$ is a projection operator onto states of fixed center-of-mass energy, $E$.
The probability is unity unless the initial state is projected
also onto a subspace which does not commute with the Hamiltonian;
a projection operator $P_Q$ onto states of fixed winding number $Q$
is implicit in our choice of a classical field with this property.
Furthermore, the inclusive sum is over all final states built above
a neighboring sector of the periodic vacuum.

This quantity is relevant to the study of multiparticle cross-sections
because, when summed over all initial states,
\beq
\label{eq:microcanonical}
\sigma_E \: = \: \sum_a\:\sigma_E \left( \{ \ak \} \right)
\eeq
it gives the ``microcanonical'' transition probability in the one-instanton
sector;
the probabilities of transition from all states of energy $E$ are equally
weighted in this sum.
When evaluated in the saddle point approximation,
$\sigma_E $ yields the maximal transition probability among all states with
energy $E$.
It sets therefore an upper bound on the two-particle inclusive cross-section
in the one-instanton sector \cite{KRT3}.

By writing (\ref{eq:probability}) as a path integral of an
exponential,
its saddle point nature in the limit $g\rightarrow 0$ can be made
evident, provided all terms in the exponent are $\sim 1/g^2$.
This is the case if both the number of particles in the initial and final
states is $O(1/g^2)$.
The complex time contour $C_T$ provides a natural way of incorporating
non-vacuum
boundary conditions at the initial and final times.
In the semiclassical approximation, the initial and final states are coherent
states of the form:
\beq
\label{cohstate}
\vert\, \{ d ({\bf k}) \} \,\rangle \: = \:
\exp\left[ \, \int d {\bf k}\, d({\bf k})\, \hat{a}^+ ({\bf k})\, \right]
\,\vert\, 0\,\rangle
\eeq
The creation operator is $\hat{a}^{+}({\bf k})$
and all color and polarization indices have been suppressed.
The complex amplitudes $d({\bf k})$ are determined by
the free-field asymptotics of the solution at the ends of the contour $C_T$
\cite{KRT,RSTnew,GP}.

In order to calculate the Fourier transforms of the gauge fields at
large Minkowski time,
we note that at large $t$ the solution (\ref{soltn}) represents
a thin shell of energy, expanding with the speed of light.
Then,
the surface energy density decreases like $1/r^2\sim 1/t^2$ and we expect
the nonlinear terms to become subdominant in the infinite time limit.
Hence, as $t\rightarrow\infty$,
the solution reduces to a solution of the free equations of motion.
The calculation of the Fourier transforms of the fields at initial time,
on the complex part of the contour $C_T$, is less straightforward,
since  at large early times the contour is trapped between two
singularities of the solution.
However, the Fourier transforms of the fields at the initial and final times
differ only by the residues of the solution (\ref{soltn}) at its poles
in the complex-$r$ plane, as a result of its analytic structure \cite{GP}.
This allows us to calculate the complex amplitudes of the initial and most
probable final state.

We find for the total average number of particles in the final state
\beq
\bar{N}_{\rm fin} \: =  \: {3\,\epsilon\,\pi^2\over g^2} \, ,
\eeq
which exactly coincides with the energy of the classical solution
 \cite{LUS, GP}.
The saddle point conditions arising from the integration over the initial
values of the fields  determine the initial coherent state
in terms of the asymptotics of the solution  \cite{RT}.
The average number of particles  in the initial state is found
to be \cite{GP}:
$$
\bar{N}_{\rm in}\: \sim\: \epsilon^{1/7}\,\bar{N}_{\rm fin}.
$$
Our solution describes therefore a transition from a state with a
smaller number of particles, $\bar{N}_{\rm in} $,
to a state with a larger number of particles, $\bar{N}_{\rm fin} $,
their ratio being controlled by the small parameter,
\mbox{$\epsilon $ \cite{RSTnew,GP}.}

However,
our solution does not maximize the microcanonical transition probability
(\ref{eq:microcanonical}).
It does not give the maximum transition probability at a given energy.
Thus, it can not be used to provide an upper bound on the $2\rightarrow n$
process cross-section.

We have found that a subclass of the SO(4)-conformally invariant
solutions found by L\"uscher and Schechter exhibits a number of remarkable
properties on a suitably chosen complex time contour:

\begin{enumerate}
\item The semiclassical suppression is equal to the action of the
BPST instanton, \mbox{${\rm Im} \,  S = \once$.}
This quantity controls the semiclassical exponential
dependence of a transition probability between coherent states.
\item The topological charge of the solution is equal to
the BPST instanton charge, \mbox{$Q = 1$.}
Thus,
the solution may have a direct interpretation for
fermion number violating processes \cite{FKS}.
\item It solves the boundary value problem for the transition
probability in the one-instanton sector from a coherent state
with a smaller number of particles, to a state with a larger number
of particles. This property makes the solution interesting for
the investigation of $2\rightarrow n$ processes with fermion number violation
at high energies \cite{RT}.
\end{enumerate}

Thus,
the L\"{u}scher-Schechter solution considered in this paper provides
an analytical benchmark for future numerical computations of
many particle transition amplitudes in Yang-Mills theory.

The assumption of conformal symmetry may allow a straightforward extension
of the ideas presented here to a few more complicated field equations,
coupled to the Yang-Mills equations.
Minkowski time solutions of the field equations for a scalar triplet
and fermion fields coupled to gauge fields have already
appeared in the literature \cite{DON}.
It may be interesting to investigate the properties of these solutions
on the complex time contour,
with an eye towards incorporating the additional fields of the Standard
Model in this formalism.
In particular, it may be possible to understand the process of fermion
number or chirality violation in the Dirac-Yang-Mills system
on the complex time contour.

However,
the high degree of symmetry assumed here clearly limits the scope
of the results.
The spherical symmetry ($SO(3)_{\rm rot} \subset SO(4)_{\rm conf}$)
of the solution has led us astray from the problem of high-energy $2\rightarrow
n$ processes.
The solution in Minkowski time has the form of a spherical shell of energy,
which collapses from infinity, then, at $t=0$, bounces back and expands with
the speed of light.
Clearly,
such a classical field configuration is a poor approximation
to an initial state of two highly energetic colliding particles.
Physical intuition would lead one to believe that a solution
with only a cylindrical symmetry might be a better candidate.

The assumption of conformal symmetry has also made less transparent
an important application of this formalism: the electroweak theory.
The mass scale $v \simeq 246\, $ GeV of electroweak theory explicitly
breaks the classical conformal invariance of the pure gauge theory.
In the case that the center-of-mass energy $E$ greatly exceeds the
scale of symmetry breaking $v$,
the Yang-Mills theory considered here may correctly describe the
classical behavior of the gauge sector of the electroweak theory.
Then, the results of this paper have direct relevance to the behavior
in this energy region \cite{RSTnew,GP}.

The complex-time solution presented here may also be considered
the ``core'' of a constrained solution in a spontaneously-broken
gauge theory \cite{AFF}.
In the Euclidean approach,
there are {\it no} exact finite-action solutions to the electroweak gauge-Higgs
field equations.
In this case,
the constrained expansion is a device to obtain the {\it approximate} solutions
which provide the dominant semiclassical contribution to scattering amplitudes.
In the complex time approach however,
nothing prevents the existence of an {\it exact} solution to the Minkowski
gauge-Higgs equations,
which would represent an additional saddle point contributions to some
transition amplitudes.
The effect of symmetry-breaking on the solution presented in this paper
has yet to be explored.

\end{document}